\documentclass[apj,numberedappendix]{emulateapj}
\usepackage{graphics,epsf}
\usepackage{amsmath}                
\usepackage{amsfonts}               
\usepackage{amssymb}                
\usepackage{epsfig}                 
\usepackage{graphicx}
\usepackage{float}
\usepackage{color}
\usepackage[para,online,flushleft]{threeparttable}

\newcommand{\km}{{~\rm km}}
\newcommand{\s}{{~\rm s}}

\newcommand{\erg}{{~\rm erg}}
\newcommand{\yr}{{~\rm yr}}

\newcommand{\nar}{{~\rm New Astronomy Reviews}}

\newcommand{\pasa}{{~\rm Publications of the Astronomical Society of Australia}}

\begin{document}

\title{The class of supernova progenitors that result from fatal common envelope evolution}
\author{Noam Soker\altaffilmark{1,2}}

\altaffiltext{1}{Department of Physics, Technion -- Israel Institute of Technology, Haifa 32000, Israel; soker@physics.technion.ac.il}
\altaffiltext{2}{Guangdong Technion Israel Institute of Technology, Shantou, Guangdong Province 515069, China}

\begin{abstract}
I construct the class of supernovae and supernova progenitors that result from fatal common envelope evolution (CEE). The fatal CEE progenitors are stellar binary systems where a companion spirals-in inside the envelope of a giant star and merges with the core. The companion can be a neutron star (NS; or a black hole) that destroys the core and by that forms a common envelope jets supernova (CEJSN), a white dwarf (WD) that merges with the core to form a massive WD that later might explode as a Type Ia supernova (the core degenerate scenario), or a main sequence companion. In the latter case the outcome might be a core collapse supernova (CCSN) of a blue giant, a CCSN of type IIb or of type Ib. In another member of this class two giant stars merge and the two cores spiral-in toward each other to form a massive core that later explodes as a CCSN with a massive circumstellar matter (CSM). I discuss the members of this class, their characteristics, and their common properties. {{{{ I find that fatal CEE events account for $\approx 6-10 \%$ of all CCSNe, }}}} and raise the possibility that a large fraction of peculiar and rare supernovae result from the fatal CEE. The study of these supernova progenitors as a class will bring insights on other types of supernova progenitors, as well as on the outcome of the CEE. 
\end{abstract}


\section{INTRODUCTION}
\label{sec:intro}
 
Although recent studies of the common envelope evolution (CEE) reveal many properties of the binary interaction and its outcomes (e.g., \citealt{Passyetal2012, RickerTaam2012, Ohlmannetal2016, Staffetal2016MN8, NandezIvanova2016, Kuruwitaetal2016, Iaconietal2017, Chamandyetal2018, GlanzPerets2018, Iaconietal2018, MacLeodetal2018a, MacLeodetal2018b,  Chamandyetal2019, MichaelyPerets2019, Reichardtetal2019, WilsonNordhaus2019, IaconiDeMarco2019}), there are still many open questions. Two open questions are how to calculate the final orbital separation and at what conditions the companion launches jets.

Traditionally (e.g., \citealt{Ivanovaetal2013} for a review), CEE studies have been looking for the final orbital separation between the core of the giant star and the more compact companion by considering the channelling of the orbital energy of the core-companion binary system to the removal of the common envelope. It seems that in many cases the final outcome will be the merger of the companion with the core (e.g., \citealt{DeMarcoIzzard2017} for a review), in what I term a fatal CEE. As well, in many cases the companion launches jets as it accretes mass from the envelope through an accretion disk, and the jets might contribute a non-negligible amount of energy to envelope removal. The companion might launch jets as it grazes the envelope (e.g., for hydrodynamical simulations,  \citealt{Shiberetal2017, Shiber2018}) and later during the CEE itself if it enters a CEE phase (e.g., \citealt{BlackmanLucchini2014}; \citealt{Shiberetal2019} for hydrodynamical simulations). Later, the companion might launch jets as it exists the CEE (e.g., \citealt{Soker2019CEexit}). 

The exploration of the CEE requires us to understand the observational consequences of all CEE channels (see review by, e.g., \citealt{DeMarcoIzzard2017}), including those of core-companion merger and cases where the companion launches jets (e.g., \citealt{Soker2016Rev} for a review). 
The observational manifestations of companion-core merger and jets' launching in cases of giant stars in the mass range $\simeq 1-8 M_\odot$ are related to the formation of planetary nebulae. 
The merger of a companion, {{{{ a planet, a brown dwarf, a main sequence star or a white dwarf (WD), }}}} with the envelope of a small primary star (not a fully developed giant) releases a large amount of energy that can result in an intermediate luminous optical transient (ILOT; e.g., \citealt{RetterMarom2003, Tylendaetal2011, Nandezetal2014, Ivanova2017}). {{{{ When the primary is a giant star the onset of the CEE by itself might lead to an ILOT radiation that comes mainly from recombination \citep{Ivanovaetal2013Science}. }}}} The merger of the companion with the core can lead to an ILOT as \cite{Tylendaetal2013} suggest for the ILOT OGLE-2002-BLG-360. 

{{{{ I here note that there are alternative names to ILOTs.
\cite{Jencsonetal2019}, for example, use intermediate luminosity red transients for explosions of extreme asymptotic giant branch stars, and luminous red novae for merging stars. They do not use the name ILOT at all. 
I here follow \cite{KashiSoker2016RAA} and include all these transients and those of luminous blue variables and pre-CCSN eruptions under the name ILOTs.   }}}} 

When the primary is a giant star, the onset of the CEE requires the companion to accrete mass at a high rate for the event to be an ILOT that is powered by accretion. The accretion process leads to emission either directly or by the collision of jets that the companion launches with the envelope and the wind from the giant star. This event appears as an ILOT during the formation of some bipolar planetary nebulae \citep{SokerKashi2012}.
Jets that the companion launches in binary systems, including those that experience the CEE, might therefore form a bipolar nebula, as studies have been suggesting for many planetary nebulae (e.g.,  \citealt{Corradietal2011aa, Miszalskietal2011,Velazquezetal2011ApJ, Boffinetal2012, Tocknelletal2014, Akarasetal2015, Chenetal2016, Sahaietal2016, Sahaietal2017, JonesBoffin2017, Sowickaetal2017, Dopitaetal2018, Jones2019, Jonesetal2018, Miszalskietal2018}).

In this study I point out some observational manifestations of companion core merger during the CEE in cases where the primary giant star is more massive than in the case of planetary nebula progenitors, and/or the companion star is more compact, i.e., a WD, a neutron star (NS), or a black hole (BH). Specifically, the subject of this study is the class of binary (and triple) stellar systems that experience the fatal CEE and lead to supernovae or progenitors of supernovae. These cases are rare in the case of massive stars, adding up to about several percents of all exploding massive stars.

Many papers study the formation of Type Ia supernovae (SNe Ia; e.g., \citealt{Maozetal2014, LivioMazzali2018, Soker2018Rev, Wang2018, RuizLapuente2019}, for reviews from the past five years) and core collapse supernovae (CCSNe; for a small number of papers from the last five years see, e.g., \citealt{DeMinketal2014, Justhametal2014, SmithN2014, Taurisetal2015, vandenHeuveletal2017,  Eldridgeetal2018, VignaGomezetal2018}) from binary (and triple) stellar evolution. I differ from these studies in that in this study I establish the class of supernovae and supernova progenitors that result from the fatal CEE. 

As the field of supernovae has many open questions of its own, one of my goals in combining these two topics that have many open questions is that the two research areas of supernovae and CEE will shed light on each other. 
In section \ref{sec:Properties} I list the common properties of the members of the class, and in section \ref{sec:classmembers} I describe each of the class members that I study here. I present my summary in section \ref{sec:summary}. 

\section{Common properties}
\label{sec:Properties}
 
There are several common properties of fatal CEE supernovae, although each property by itself is  not unique to this group of supernovae. 
 
(1) \textit{A progenitor with a massive envelope.} To cause merger of a compact companion with the core the giant's envelope mass $M_{\rm env}$ must be much larger than the mass of the companion $M_2$. For example, in the case of a WD companion the ratio of envelope mass to the WD companion mass for a merger to take place should be $M_{\rm env}/ M_{\rm WD,2} \ga 4$  \citep{Sokeretal2013PTF11kx, Canalsetal2018}. 
When the companion is more likely to launch jets during the  CEE, i.e., for main sequence companions and more so for NSs, the mass ratio should be somewhat larger. I therefore estimate the condition to cause merger to be 
$M_{\rm env} \ga (4-5) M_2$.
In the case of the merger of two cores of two giants stars, at least one envelope is much more massive than the core of the second giant. 
A massive envelope is not a unique property of fatal CEE supernovae as many other CCSNe come from progenitors with massive envelopes. 
   
(2) \textit{Non-spherical circumstellar matter (CSM).} The CSM is highly asymmetrical. When the companion is a compact object that is likely to launch jets, mainly main sequence stars and NSs, the CSM will be bipolar, e.g., having two opposite lobes. Such an example is SN 1987A, where the progenitor is likely to have gone a fatal CEE with a main sequence companion (e.g., \citealt{ChevalierSoker1989, Podsiadlowskietal1990, MenonHeger2017, Urushibataetal2018}). For WD companions the outcome will be an elliptically shaped CSM, possibly with Ears (e.g., \citealt{TsebrenkoSoker2015}; `ears' refer to two opposite protrusions from the main nebula). The merger of two giants might lead to an elliptical CSM with a dense equatorial slow outflow.
The bipolar CSM morphology is not a unique property to fatal CEE because binary companions that are close to the giant progenitor but avoid CEE, as well as binaries that experience the CEE but survive, can also shape a massive bipolar CSM. 
 
(3) \textit{A Singly exploding star.} At the time of explosion itself there is a single star (or a far-away star if the system was multiple-stellar system). In the case of the core degenerate (CD) scenario, the explosion leaves no remnant. In the other cases we study here the remnant is a NS or a BH. This is not a unique property because other single star can also explode as CCSNe. 

(4) \textit{Possible strong ejecta-CSM interaction.} If the explosion occurs within a time frame of hundreds to thousands of years post-merger there might be a strong ejecta-CSM interaction that affects the light curve days to years post-explosion. In particular, this interaction can lead to superluminous CCSNe. In that respect, the general classification of superluminous CCSNe and the quest to understand them (e.g., \citealt{GalYam2019}) can benefit from the present study. The strong ejecta-CSM interaction is not a unique property because binary companions that are close to the giant progenitor but avoid CEE, as well as binaries that experience the CEE but survive, can also eject a large fraction of the envelope (and in a bipolar geometry). 

(5) \textit{Very energetic transient during the CEE or later during the core-companion merger.} There are two sources of energy that can in principle form an energetic transient event before the explosion itself. The first is the accretion of mass by the companion as it enters the envelope and spirals inside it. The companion might launch jets that when interact with the envelope lead to a bright visible event, an ILOT (e.g., \citealt {KashiSoker2018}). Such an ILOT event can take place even when the companion is outside the envelope (e.g., \citealt{SokerKashi2016}). 
The other energy source is the spiralling-in of the core-companion system into each other until they merge (e.g., \citealt{Tylendaetal2013}), including the recombination energy of the outflowing envelope. The amount of energy that each process liberates differs from one fatal CEE system to another as I describe in section \ref{sec:classmembers}. As well, the fraction of the energy that radiation carries can vary tremendously between different cases.

Although each of he above properties is not unique to supernovae of fatal CEE progenitors, the combination of several, or even all, properties in one supernova (as in SN 1987A) points to a fatal CEE progenitor. I turn now to describe the members of the class.  

\section{The class members}
\label{sec:classmembers}

\begin{figure*}
\begin{center}
\vspace*{-3.50cm}
\hspace*{-1.7cm}
\includegraphics[width=1.1\textwidth]{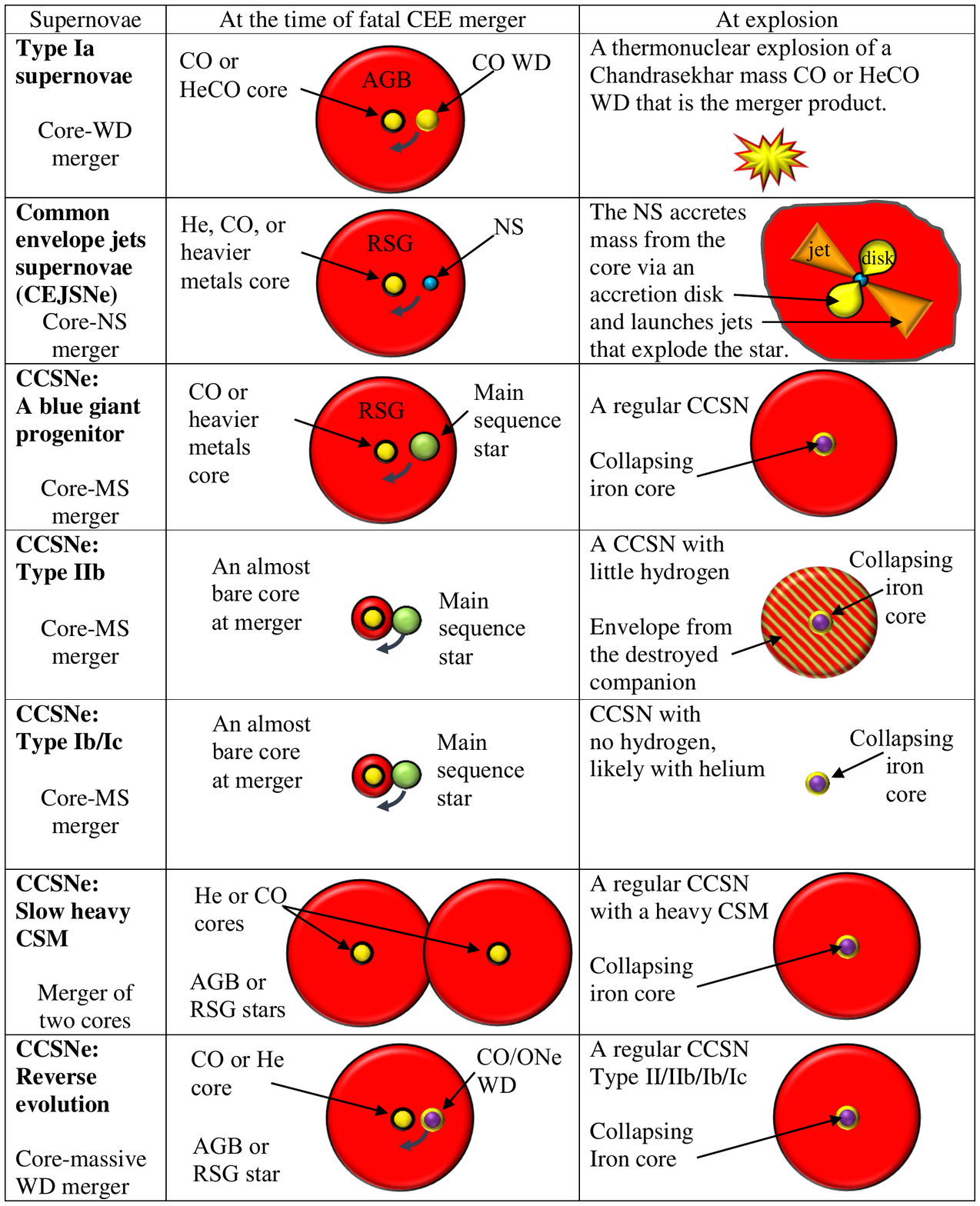}
\vspace*{-3.2cm}
\caption{The seven fatal CEE scenarios that lead to SNe that I study here. The left column gives the name of the scenario, the middle column shows the configuration when the companion merges with the core of the giant star, and the right column presents the merger product at explosion. Drawing are not to scale. Table \ref{tab:Table1} lists more qualitative and quantitative  details. 
Abbreviation: AGB: asymptotic giant branch star; CCSN: core collapse supernova; CSM: circumstellar matter; MS: main sequence star; RSG: red supergiant star. }
\label{fig:Schematic}
\end{center}
\end{figure*}

{{{{ In Fig. \ref{fig:Schematic} I schematically present the seven scenarios that I discuss here. }}}}
I list these seven fatal CEE progenitors in Table \ref{tab:Table1}, with the main properties of each progenitor. I discuss each one of these in more details in the following subsections. In the third row from the bottom I list some possible examples for the proposed scenario. The fatal CEE is not the sole scenario for these systems, and so in the second row from the bottom I list some references to studies that propose alternative scenarios. 
\begin{table*}
\tiny
\scriptsize
\begin{center}
  \caption{Supernovae from fatal CEE}
    \begin{tabular}{| p{1.6cm} | p{1.9cm}| p{1.9cm}| p{1.9cm}| p{1.9cm} | p{1.9cm} | p{1.9cm} | p{1.9cm}|}
\hline  
 Supernova type & Type Ia supernovae &  {Common envelope jets supernovae (CEJSNe)} & CCSNe: a blue giant progenitor & CCSNe: Type IIb & CCSNe: Type Ib/Ic & CCSNe: Slow-heavy CSM  & CCSNe: Reverse evolution leading to SNe II/IIb/Ib/Ic\\
\hline  
\hline  
 Initial primary mass $(M_\odot)$ & $4\la M_{\rm Z,1} \la 10$ &  $ 10 \la M_{\rm Z,1}$ & $8 \la M_{\rm Z,1}$ & $8 \la M_{\rm Z,1} \la 15$ & $10 \la M_{\rm Z,1}$  & $5 \la M_{\rm Z,1}$ &   $4 \la M_{\rm Z,1} < M_{\rm Z,2}$, but at CEE $M_{1} \ga 8.5$  \\
 
\hline  

 Primary core at merger &  CO core or hybrid HeCO core &  Helium or heavier metals & CO or heavier & Helium or heavier metals & Helium or heavier & CO or He core & CO or He core          \\

\hline  
 Companion $(M_\odot)$ & CO white dwarf &  Neutron star & main sequence $1 \la M_2 \la 3-5$ & Low mass main sequence $0.5 \la M_2 \la 1$ & Low mass main sequence $M_2 \approx 0.5 M_\odot$& A giant with $M_2 \simeq M_1$, majority with He core & CO/ONe WD  of initially more massive $5.5 \la M_{\rm Z,2} < 8.5 $          \\

\hline  
ILOT Energy (erg) & $E_{\rm I,jet} \ll E_{\rm I,mer}$; $E_{\rm I,mer} \approx 10^{50} $   &      $E_{\rm I,jet} \approx 10^{50-52}$; $E_{\rm I,mer} \approx 10^{50} $ & $E_{\rm I,jet} \approx 10^{46-47}$; $E_{\rm I,mer} \ll 10^{49}$    & $E_{\rm I,jet} \approx 10^{45-46}$; $E_{\rm I,mer} \ll 10^{48}$   &   $E_{\rm I,jet} \approx 10^{45-46}$; $E_{\rm I,mer} \ll 10^{48}$   & $E_{\rm I,jet} =0$; $E_{\rm I,mer} \approx 10^{49-50}$   &  $E_{\rm I,jet} \ll E_{\rm I,mer}$; $E_{\rm I,mer} \approx 10^{50} $        \\

\hline  
Explosion mechanism     &  Thermonuclear: $\simeq 1.4 M_\odot$ WD  & NS accretes core: fixed-axis jets& Core collapse:  jittering jets & Core collapse:  jittering jets& Core collapse: jittering jets around a preferred axis& Core collapse: jittering jets around a preferred axis & Core collapse: jittering jets around a preferred axis \\

\hline  

Explosion energy (erg)  &$E_{\rm exp} \simeq 10^{51}$& $E_{\rm exp} \approx 10^{51-53}$&  $E_{\rm exp} \simeq 10^{51}$ &  $E_{\rm exp} \simeq 10^{51}$  & $E_{\rm exp} \ga 10^{51}$      &  $E_{\rm exp} \ga 10^{51}$  & $E_{\rm exp} \ga 10^{51}$         \\
\hline  
CSM at explosion    & Elliptical with ears; in most cases gone at explosion & Bipolar; present at explosion & Bipolar: might be present, e.g., SN~1987A & Bipolar: might be present &  Bipolar: might be present   &  Elliptical: Likely to be present &   Elliptical with ears; might be present        \\
     
\hline  
Nucleo-synthesis & Products of a Chandrasekhar mass WD explosion, e.g., $^{55}$Mn &  $\approx 10 \%$ of CEJSNe produce heavy r-process elements    & Typical CCSNe    &   Typical CCSNe  & Typical CCSNe     & Typical CCSNe  &   Typical CCSNe         \\
\hline  

Fraction  & ${\rm several} \times 10 \%$ of all SNe Ia; most or all Chandrasekhar mass [S18]  & ${\rm few} \times 0.1\%$ of all CCSNe [based on GS19]& $\approx 2-3 \%$ of all CCSNe [based on Sh17]& $\approx 10-30 \%$ of SNe IIb, or $\approx 1-3\%$ of all CCSNe [here: \S\ref{subsec:SNIIb}]& $\approx 10 \%$ of SNe Ib, or $\approx 1\%$ of all CCSNe; $0-1 \%$ of SNe Ic  [here: \S\ref{subsec:SNIbc}] & $ \approx 1 \%$ of all CCSNe [based on VG19]&  $\approx 2 \%$ of all CCSNe [Sa14]      \\

\hline  
Possible examples & PTF~11kx [S13]     & iPTF14hls [SG18]; AT2018cow [S19] & SN~1987a [P90, MH17,U18] & Single stars exploding as SN IIb, possibly with bipolar CSM   &  Single stars exploding as SN Ib, possibly with bipolar CSM   &  SNe IIb/Ib/Ic + possible dense equatorial but no fast polar outflow &  CCSNe occurring 50 - 200 Myr after stellar birth [Z17]        \\

\hline  

Possible alternatives to fatal CEE & PTF~11kx from single degenerate scenario [D12]& TDE, magnetar, etc. (iPTF14hls:  List1; AT2018cow: List2) & Single star models can form blue progenitors, but not 1987A [MHD17] & Single-stars [Sra1819], but not if there is a bipolar CSM; other binary channels & Single [Y17] + some other binary scenarios [Y15], e.g., with Roche lobe overflow &  Some other binary scenarios, e.g., with Roche lobe overflow &  Merging of initially more massive star with its companion [Z17]        \\

\hline  
Some possible outcomes when the companion is in a tight binary with star `Ter' 
& Ter=MSS: CD scenario leaving a MSS; Ter=CO WD: DD scenario inside the envelope 
& Ter=NS: CEJSN + gravitational waves source
& Ter=MSS: A messy CSM (i.e., a nebula missing any kind of symmetry) 
& Ter=MSS: A messy CSM and/or a MSS bound or escaping the progenitor
& Ter=MSS: A messy CSM and/or a MSS bound or escaping the progenitor   
& Not relevant & Not likely       \\
 \hline  
%
     \end{tabular}
  \label{tab:Table1}\\
\end{center}
\begin{flushleft}
\small The different fatal CEE progenitors as the text explains. The third row from the bottom lists some possible examples, and the next to last row lists possible alternative explanations. The last row lists some possible outcomes when the companion is in a tight binary system. 
\newline
Legend: TDE: tidal disruption event; $M_{\rm Z} \equiv M_{\rm ZAMS}$ zero age main sequence mass; $E_{\rm I,jet}$ is the energy that jets that the companion launches might carry, while $E_{\rm I,mer}$ is the energy that the companion-core merger process liberates.  MSS: main sequence star. DD scenario: The double degenerate scenario for SNe Ia. Ter: The tertiary star in the system
\newline
References: List1: \cite{Arcavietal2017, AndrewsSmith2018, Chugai2018, Dessart2018, Wangetal2018, Quataertetal2019, Woosley2018}
List2: \cite{Fangetal2019, Kuinetal2019, LyutikovToonen2019, Marguttietal2019, Perleyetal2019, Quataertetal2019}; 
D12: {Dildayetal2012}; GS19: \cite{GrichenerSoker2019}; MH17: \cite{MenonHeger2017}; MHD17: For a list of single star models that might form blue CCSN progenitors see \cite{MenonHeger2017}; Sa14: \cite{SabachSoker2014}; S13: \cite{Sokeretal2013PTF11kx}; SG18: \cite{SokerGilkis2018}; S18: \cite{Soker2018Rev}; S19:  \cite{Sokeretal2019AT2018cow}; Sh17:  \cite{Shivversetal2017}; Sr18: \cite{Sravanetal2018}; Sra1819: \cite{Sravanetal2018, Sravanetal2019}; VG19: \cite{VignaGomezetal2019}; U18: \cite{Urushibataetal2018}; Y17: \cite{Yoon2017}; Y15: \cite{Yoon2015}; Z17: \cite{Zapartasetal2017} .  
\end{flushleft}
\end{table*}

Many binary systems are actually triple-stellar systems. I limit the discussion here to include only triple stellar evolutionary routes that include a tight (short) binary systems that enters the common envelope. \cite{SabachSoker2015} discuss some possible outcomes of such an evolution {{{{ to explain the formation of the pulsar with two white dwarfs triple system PSR~J0337+1715 }}}}, and 
\cite{Antonietal2019} conduct numerical simulations of mass accretion by a binary system in a homogeneous medium like the interstellar medium.  
The outcome might be that the two stars of the tight binary system merge, or that only one of them merges with the core and the other is ejected from the system or stays bound. In briefly speculating on possible outcomes of these rare cases of triple stellar evolutionary routes I will follow the results of \cite{SabachSoker2015} and \cite{Antonietal2019}.

\subsection{The core degenerate scenario of SNe Ia}
\label{subsec:CoreDegenerate}
 
There are five different binary scenarios to explain the formation of progenitors of SNe Ia, from which four scenarios involve the CEE (see recent reviews by \citealt{LivioMazzali2018, Soker2018Rev, Wang2018, RuizLapuente2019}). Only the core degenerate (CD) scenario involves a fatal CEE, in this case where a WD companion merges with the core {{{{ (for more on the general properties of the CD scenario see, e.g., \citealt{Soker2018Rev} and references therein). }}}} There are evidences that SNe Ia involve both Chandrasekhar mass exploding WDs, i.e., $M_{\rm Ch}$ explosions, and sub-$M_{\rm Ch}$ explosions. For example, some isotopes, such as $^{55}$Mn, require that a non-negligible fractions of SNe Ia comes from $M_{\rm Ch}$ explosions (e.g., \citealt{Seitenzahletal2013, Bravo2019}). 
It is possible that the CD scenario is the main scenario responsible for  $M_{\rm Ch}$ explosions (e.g., \citealt{TsebrenkoSoker2015, Soker2018Rev}). {{{{ By $M_{\rm Ch}$ I refer to exploding WDs in the mass range $\simeq 1.3-1.5 M_\odot$. I note that even if the combined mass of the WD and core that merge is much larger than that range, the CD scenario assumes that in most cases the merger processes self-regulates itself to end with a merger product with a mass in that range   }}}}
 
The thermonuclear explosion might take place already during the CEE, or very shortly thereafter \citep{LivioRiess2003}. \cite{LivioRiess2003} considered the WD-core merger to be a rare case of thermonuclear events that occurs during or shortly after the CEE. The CD scenario assumes the merger to be more common and that the explosion might occur as well a long time after the CEE (e.g., \citealt{Soker2018Rev}).
 
The CD scenario differs from the other members of the fatal CEE progenitor class in two  qualitative ways. (1) Nuclear energy powers the explosion, rather than gravitational energy in the other supernovae. (2) The WD cannot accrete mass at even a moderate mass accretion rate because the WD develops nuclear burning on its surface that causes its envelope to expand (e.g., \citealt{Hachisuetal1999}). This implies that the WD does not launch energetic enough jets {{{{ (weak jets are possible, e.g., \citealt{TsebrenkoSoker2015}). }}}} Only the in-spiral process powers the pre-explosion ILOT. The energy that the merger liberates to the envelope is 
\begin{eqnarray}
\begin{aligned}
E_{\rm I,mer} & \simeq \frac{1}{2} \frac {G M_{\rm core} M_2}{R_{\rm core} + R_2}
\approx 10^{50}
\\ & \times 
\left( \frac{M_{\rm core} M_2}{M^2_\odot} \right)
\left( \frac{{R_{\rm core} + R_2}}{0.02R_\odot} \right)^{-1} \erg, 
\end{aligned}
\label{eq:EImer}
\end{eqnarray}
where $M_2$ and $R_2$, and $M_{\rm core}$ and $R_{\rm core}$ are the mass and radius of the companion and core, respectively. 
I list the ILOT energy sources in Table \ref{tab:Table1}, together with the other relevant properties of the  CD scenario. 

There are other outcomes of a fatal CEE of a WD inside a giant star (e.g., \citealt{SabachSoker2014, Zapartasetal2017}), some of which I discuss in section \ref{subsec:reverse}.  

{{{{ In a recent review paper \citep{Soker2018Rev} I estimated the fraction of SNe Ia that come from the CD scenario to be about 60 per cent of all SNe Ia. The rest SNe Ia come mainly from the DD scenario. But the uncertainties are large, and in light of some recent studies of the DD scenario I estimate that each of these two scenarios account for about half of all CCSNe.  }}}}

In rare cases the WD that enters the envelop will be part of a tight binary system. If the tertiary star is a main sequence star it might be ejected from the system or might stay bound after the WD merges with the core. The outcome might be a SN Ia in the CD scenario but with a surviving main sequence star. In even rarer cases the tertiary star might be a WD and the explosion might leave a surviving WD. 
 
Another very speculative route is the one where the two WDs in the tight binary system merge and then might lead to a SN Ia through the double degenerate (DD) scenario inside a common envelope. This will leave the bare core which might evolve to a helium WD (sdO star) if the giant was on the red giant branch, or to a CO WD if the giant was on the  asymptotic giant branch. 

\subsection{Common envelope jets supernovae (CEJSNe)}
\label{subsec:CEJSNe}

In CEJSNe a NS companion spirals all the way to the core and destroys it by tidal interaction and by accreting mass and launching jets.  
The CEJSN scenario is possible because neutrino cooling allows NSs to accrete mass at very high rates of $\dot M_{\rm acc} \ga 10^{-3} M_\odot \yr^{-1}$ \citep{HouckChevalier1991, Chevalier2012}. This accretion is very likely to occur via an accretion disk \citep{ArmitageLivio2000, Papishetal2015, SokerGilkis2018}, and the CEJSN scenario assumes that this accretion disk launches jets. A BH companion can launch jets in the same manner. When the NS (or BH) accretes mass from the core the mass accretion rate can be very high and within a short timescale of minutes to tens of minutes. The outcome is an explosion that mimics a CCSN and even a super-energetic CCSN with energies up to $\approx 10^{53} \erg$. 
{{{{ I base the claim that CEJSN explosions mimic CCSN explosions on the assumption that most CCSNe are driven by jets, either jets with varying directions in the jittering jets explosion mechanism, or jets with a more or less constant symmetry axis in cases of rapid pre-collapse core rotations
(e.g., \citealt{Soker2017RAA}). }}}} 

In the extreme cases, when the NS accretes more than about $1 M_\odot$, it might collapse into a BH. In these cases the explosion energy is $E_{\rm exp} \ga {\rm few} \times 10^{52} \erg$ \citep{SokerGilkis2018}. 
 
A very energetic ILOT is expected to precede the explosion itself with energies of up to $10^{52} \erg$ (hence, literally, it is not an `intermediate' anymore), although most of it results in kinetic energy rather than radiation (e.g., \citealt{SokerGilkis2018}). 
{{{{ Most of the energy of the ILOT comes from the jets that carry a kinetic energy of 
\begin{equation}
E_{\rm jets} 
= 10^{52}
\left( \frac{M_{\rm jets}}{0.1 M_\odot} \right)
\left( \frac{v_j}{10^5 \km \s^{-1}} \right)^{2} \erg, 
\label{eq:Ejets}
\end{equation}
where $M_{\rm jets}$ is the mas in the two jets and $v_j$ is the initial velocity of the jets that is about equal to the escape velocity from the NS. To launch a mass of $0.1 M_\odot$ the NS should accrete a mass of 
$\simeq 0.5-1 M_\odot$, that might turn it to a BH. For energies much below $10^{52} \erg$ the NS stays a NS even after accretion. 
}}}}
The gravitational energy of the core-NS system is small in these cases. I list the ILOT energy sources in Table \ref{tab:Table1}. 
{{{{ \cite{SokerGilkis2018} attribute the 1954 pre-explosion outburst of iPTF14hls to an ILOT event where the NS accreted mass from the envelope while still being on an eccentric orbit grazing the envelope. }}}}

\cite{Sokeretal2019AT2018cow} list the different CEJSN channels. These include also a channel that leads to r-process nucleosynthesis in the jets, the CEJSN r-process scenario \citep{GrichenerSoker2019}. As well, in some cases when the early evolution removes all hydrogen and possibly also helium from the envelope, the CEJSN can lead to a stripped-envelope CEJSN channel. 

{{{{ \cite{GrichenerSoker2019} estimate that the rate of CEJSN r-process events is $\approx 0.001$ times that of all CCSNe. Considering the other CEJSN scenarios, I crudely estimate that all CEJSN events amount to ${\rm few} \times 0.001$ of the total rate of CCSNe. }}}}

{{{{ The merger of the NS with the core leads to emission of gravitational waves with unique signatures, as \cite{Nazinetal1997} noted for a similar process. In a recent detailed study \cite{Ginatetal2019} suggest that next-generation space-based gravitational wave detectors will be able to detect gravitational waves from NS-core merger. }}}}

In cases where the NS (or BH) is part of a NS-NS (or NS-BH or BH-BH) binary system, the drag inside the envelope, and more so inside the destroyed core, might cause the two NSs to merge inside the giant. The outcome is a CEJSN that is also a gravitational waves source where the spiralling-in to merger is driven also by drag rather than by gravitational waves alone.  {{{{ As emission of gravitational waves drives the final merger of the tight binary system, the properties of the gravitational waves in the last several seconds are as in regular merger NS-NS or NS-BH or BH-BH binary systems. }}}}

\subsection{A blue giant progenitor}
\label{subsec:Bluegiant}

The motivation to develop a fatal CEE scenario for SN~1987A came shortly after its explosion to account for its asymmetrical explosion \citep{ChevalierSoker1989} and to its blue giant progenitor \citep{Podsiadlowskietal1990}, with follow-up studies, e.g., \cite{MenonHeger2017} and \cite{Urushibataetal2018}. 

In this scenario the companion {{{{ that merges with the  core }}}} is a main sequence star. {{{{ Before the companion merges with the core }}}} it can accrete mass and then launch jets to shape the CSM, possibly the outer rings of the CSM of SN~1987A. The expected energy that the jets carry in this case is $E_{\rm I,jet} \approx 10^{46}- {\rm few} \times 10^{47} \erg$ for an accreted mass during the entire CEE of $\approx 0.1-1 M_\odot$ and for cases where about $10 \%$ of the accreted mass is carried by the jets at the escape speed from the main sequence star. As the companion spirals-in inside the envelope it releases gravitational energy that might lead to an equatorial outflow and enhanced mass loss rate. 
  
The destruction of the main sequence star at $\approx 2 R_\odot$ near the core releases gravitational energy of $\approx 10^{49} \erg$ for $M_{\rm core} {M_2} \simeq 10 M^2_\odot$ at the final CEE (by equation \ref{eq:EImer}). Because of the still massive envelope, a large fraction of this energy goes to inflate  the mass of the destroyed companion back into a giant envelope. Only a small fraction of this energy is left to power an ILOT, and so I take $E_{\rm I,mer} \ll 10^{49} \erg$. When the envelope mass that is left at merger is much lower, as I discuss in coming subsections, the core-companion merger can contribute more to the kinetic energy of the outflow and to the light of the ILOT. 

The explosion itself in this scenario is of a regular CCSN. The rotation of the core is slow and so in the frame of the jets feedback mechanism for CCSNe the jets are formed from stochastic angular momentum accretion. Namely, the jittering jets explosion mechanism (e.g., \citealt{Soker2017RAA, Soker2019SASI}). 

\cite{Pastorelloetal2012} estimate that $1- 3 \%$ of all CCSNe are 1987A-like events and \cite{Shivversetal2017} estimation for this numbers is $\approx 3 \%$. The  heterogeneous properties of this class hints on more than one evolutionary channel. I very crudely estimate that the fatal CEE scenario accounts for most 1987A-like events. This amounts to $\approx 2-3 \%$ of all CCSNe. 

In cases where a tight binary system of two main sequence stars enters the envelope there are three fatal CEE outcomes. ($i$) The two main sequence stars merge inside the envelope and then the merger product merges with the core. ($ii$) One star merges with the core and the other is ejected from the envelope and either escape the system or stays bound. ($iii$) One star merges with the core and removes the entire envelope, and the second main sequence star orbits the core. 
This will lead to a SN Ib or SN Ic, which I discuss below. 

In all these triple star evolutionary routes the ejected nebula might be `messy', i.e., highly asymmetrical and lacking any axisymmetrical symmetry and even lacking mirror symmetry about the equatorial plane, like studies suggest for messy planetary nebulae \citep{Soker2016messy, BearSoker2017, Hilleletal2017}. 

\subsection{Fatal CEE as progenitor of some SNe IIb}
\label{subsec:SNIIb}
 
\cite{Nomotoetal1995} suggested a fatal CEE scenario for SNe IIb, where the spiralling-in of the companion inside the envelope removes a large mass from the envelope, leaving only a very low mass hydrogen-rich envelope. Similarly, \cite{Youngetal2006} considered a fatal CEE for the progenitor of Cassiopeia~A. 
One might consider also the mass of the companion after the core destroys it. \cite{Lohevetal2019} present the following fatal CEE scenario that might lead to progenitors of some, but definitely not all, SNe IIb. According to \cite{Lohevetal2019} a main sequence companion of mass $M_2 \approx 0.5-1 M_\odot$ spirals-in inside the envelope of the giant and removes all, or most of, the hydrogen-rich envelope. The companion further spirals-in and tidal forces destroy it on to the helium or CO core of the massive giant star. The mass of the destroyed companion forms now the envelope of a single giant star. The end product is a giant star but with small amount of hydrogen. The merger itself releases energy that might expel more hydrogen. At explosion, the envelope does contain hydrogen, but an amount of $M_{\rm env} \approx 0.1-1 M_\odot$. To leave little hydrogen in the envelope the companion cannot be too massive, i.e., $M_2 \la 1 M_\odot$, and hence for the companion to remove the entire hydrogen-rich envelope the giant cannot be too massive. I crudely put an upper limit of zero age main sequence mass of $M_{\rm ZAMS,1} \la 15 M_\odot$, but this limit requires further studies.    

The energy of the ILOT in this case is somewhat lower than that in the process of the formation of a blue giant. However, due to the lower envelope mass the effect is likely to be larger, like removing more hydrogen from the leftover original envelope and from the envelope of the merger product. 

There are other scenarios for the formation of SNe IIb that involve single stars (e.g., \citealt{KotakVink2006, Sravanetal2018, Sravanetal2019}), binary interaction with Roche lobe overflow (e.g., \citealt{Claeysetal2011, Sravanetal2018, Sravanetal2019}), and the grazing envelope evolution (GEE) scenario \citep{Soker2017IIb}. I estimate that the fatal CEE scenario accounts for much less than half of all SNe IIb. 
\cite{Shivversetal2017} estimate that SNe IIb make $\approx 10\%$ of all CCSNe. I very crudely estimate that the fatal CEE scenario for SNe IIb accounts for $\approx 1-3\%$ of CCSNe. 

If the companion is in a tight binary system, the tertiary star might survive. {{{{ As well, the two stars in the tight binary system might merge and later the merger product itself merges with the core. }}}} In any case, the main effect is that the tight binary system forms a messy CSM (see section \ref{subsec:Bluegiant}).

\subsection{Fatal CEE as progenitor of some SNe Ib}
\label{subsec:SNIbc}

For a main sequence star to remove most of the hydrogen from a massive star it cannot be too light, $M_2 \la 0.5 M_\odot$. After it merges with the core its hydrogen-rich envelope forms an envelope around the core. To become a progenitor of SN Ib the star (and now there is only one star in the system) must lose all of its hydrogen, {{{{ and hence the companion cannot be too massive, i.e., $M_2 < 1 M_\odot$. For $M_2 \approx 0.5 M_\odot$ }}}} it is possible that the star will indeed lose all of its hydrogen-rich envelope {{{{ via its strong wind }}}} before it explodes as a CCSN \citep{Lohevetal2019}.

The merger of the companion with the core spins up the core. {{{{ Due to mass loss the core will slow down, but I expect that at core collapse its spin be faster than that of a core of star in a single star evolution. }}} }
In the jittering jets explosion mechanism the rapidly rotating core implies that the jets that the newly born NS launches jitter around a preferred axis which is the angular momentum axis of the pre-collapse core.
Jets with little jittering are less efficient in exploding the entire core, and this implies that the jets carry much more energy than the binding energy of the star (e.g., \citealt{Gilkisetal2016}).  
The explosion energy might be more than the canonical value, i.e., the explosion energy is $\approx 10^{51-52} \erg$.

This scenario for SNe Ib requires a much deeper study. It is hard for me now to estimate the fraction of events that result from this scenario. I only crudely estimate that the fatal CEE accounts for $\approx 10\%$ of SNe Ib, but the number can be lower. I encourage a further study of the fatal CEE channel of SNe Ib. 

The properties of the ILOT that precedes the explosion is similar to that in the fatal CEE scenario for SNe IIb. 

To become a SN Ic the star needs to remove all of its helium envelope. This is a difficult task after the companion gas becomes the envelope of the core. It is not clear if the fatal CEE scenario with a main sequence stellar companion can form a SN Ic. It might be that a fatal CEE with a NS in some cases forms a SN Ic. {{{{ Since this is a CEJSN, this scenario leads to a type Ic CEJSN. }}}} I estimate that the fatal CEE scenario accounts for $0-1 \%$ of SNe Ic.  

As SNe IIb and Ib compose each about a third of stripped envelope CCSNe \citep{Shivversetal2017}, I crudely estimate that fatal CEE evolutionary channels with a main sequence companion account for $\approx 10 \%$ of all stripped envelope CCSNe. 

The possible outcomes in cases where the companion is in a tight binary system with another main sequence star are similar to those in the fatal CEE scenario for SNe IIb.   
   
\subsection{CCSNe with a Slow-heavy CSM}
\label{subsec:SNgiants}
 
In this scenario the two stars have very close mass at their zero age main sequence evolutionary phase, and their orbital separation is such that both should become giants for them to merge.
 
After the primary more massive star expands and becomes a giant star it tidally interacts with the secondary star. The secondary star of about the same mass as the primary brings the system to synchronisation, i.e., the spin period of the primary star and the orbital period are the same. 
Mass transfer from the giant to the companion is possible. 
Before the primary ends its evolution the secondary star expands to become a giant. Now there is a strong tidal interaction of the secondary star and the primary. To bring the secondary rotation to synchronisation the orbital separation must decrease. At this phase there is a strong tidal interaction and relatively rapid rotation of both stars, leading to a relatively higher mass loss rate. This mass loss removes more angular momentum from the system, bringing the two stars to form a common envelope. This qualitative scenario requires quantitative study as we are currently conducting.  
 
As the two stars are giants, neither of them launch jets that can supply extra energy to remove the envelope. As both stars contribute to the common envelope, the envelope is quite massive, and to remove it the two cores must spiral-in quite close to each other. In some cases they will merge.

This fatal CEE has the following general properties. ($1$) Because the stars do not launch jets, the CSM does not have a component of very fast polar outflows. ($2$) The outflow is relatively massive as both stars contribute to the envelope. ($3$) The general outcome is an elliptical CSM, i.e., slower and denser equatorial outflow and somewhat faster polar outflow. ($4$) the energy of the ILOT comes mainly from the merger of the two cores, and its value is about the same as in the CD scenario for SNe Ia if both cores are CO-rich, or lower energy if one of them or both are helium cores. 
($5$) As both stars are stripped from most of their envelope, the outcome is an envelope-stripped CCSN (IIb/Ib/Ic). 
($6$) If the core of the primary star is a CO core or later, the explosion occurs shortly {{{{ (within thousands of years) }}}} after merger and ejecta-CSM interaction takes place (might lead to a superluminous CCSN). If both cores are He core, then the dense CSM might dispersed before explosion occurs {{{{ tens of thousands of years up to hundreds of thousands of years later. }}} }
 
Other types of binary interaction can lead to some of the above properties. But most cannot explain all of the above properties.  {{{{ For example, a massive main sequence stellar companion that survives the CEE can remove large amount of mass from the common envelope. Later the core collapses and explodes within a dense CSM. But such a main sequence companion might launch jets before it enters the envelope, leading to a CSM with fast polar outflows, something that is not expected in the case of two merging giants. The same holds for a case of a massive main sequence companion that does not enter the envelope, but accretes mass from the envelope via a Roche lobe overflow, and by a strong gravitational interaction removes mass from the system and forms a massive CSM. }}}} In any case, if we can infer that at explosion there is only one star in the system, then the fatal CEE scenario seems to be the most likely explanation.

One interesting channel of the fatal CEE of two giants is the possibility of two giants that cannot explode by themselves to form a massive enough merger product for a CCSN or an electron capture SN. \cite{Zapartasetal2017}, for example, already studied very similar evolutionary routes. If the two stars are in the mass range of $\simeq 4-8 M_\odot$ they will not explode without the merger. The merger can bring the core of the merger product to be massive enough to explode as a CCSNe. We are currently further studying this possibility. 

{{{{ \cite{VignaGomezetal2019} study the merger of two post main sequence very massive stars that might lead to pair instability supernovae, i.e., with zero age main sequence mass of $\ga 45 M_\odot$. In these very rare cases (that I do not list in the table) the final explosion is a pair instability supernova, rather than a CCSN. 
\cite{VignaGomezetal2019} estimate that the rate of these merger is in the range of about $10^{-5} -  0.003$ times that of the rate of CCSNe. Since the stars for the present scenario can have much lower masses, I crudely estimate that the rate of merger of two giants is $\simeq 0.01$ the rate of all CCSNe. }}}}

As both stars are giants, neither of them is in a tight binary system when they merge. Therefore, the triple stellar evolutionary route I discuss here is not relevant in this case. 

\subsection{CCSNe from reverse evolution}
\label{subsec:reverse}

In this evolutionary channel it is the initially more massive star, that here I term the secondary star, of mass $5.5 \la M_{\rm ZAMS,2} \la 8.5 M_\odot$, that forms a CO WD or an ONeMg WD that becomes the compact companion in the fatal CEE.  The primary star at the onset of the CEE was the initially less massive star of mass $4 \la M_{\rm ZAMS,1} < M_{\rm ZAMS,2} $, that accreted mass from the secondary star to become massive enough to undergo a CCSN (or an electron capture supernova which I treat here together with CCSNe).
Because the system forms the WD before it forms the NS, this is referred to as \textit{WD - NS reverse evolution}. 
  
\cite{SabachSoker2014} studied the many possible outcomes of the WD - NS reverse evolution, including some thermonuclear explosions and the formation of compact binary systems (see their figure 1 and 2). \cite{Zapartasetal2017} conducted population synthesis study of some of these evolutionary routes.
{{{{ If the thermonuclear explosion occurs when there is still hydrogen and/or helium rich envelope this is not a SN Ia. If the merger product experiences a thermonuclear explosion after it loses all if its hydrogen and helium, then this is a sub-channel of the CD scenario. }}}}
Here I focus on WD-NS reverse evolutionary routes that lead to a fatal CEE that ends with a CCSN (including an electron capture supernova). 
 
The merger of the WD with the core forms a rapidly rotating more massive core, that eventually collapses to form a NS. {{{{ I consider the case where the newly born NS launches jets that explode the star. }}}} In the jittering jets explosion mechanism the rapidly rotating core implies that the jets that the newly born NS launches jitter around a preferred axis which is the angular momentum axis of the pre-collapse core. The explosion energy might be more than the canonical value, i.e., $E_{\rm exp} \ga 10^{51} \erg$. 

Another possible outcome is that a merger of an ONeMg WD companion with the core ends in accretion-induced collapse (e.g., \citealt{SabachSoker2014}). The newly born NS accretes mass and launches jets that explode the star. The observational consequences are similar to those of CCSNe. 

\cite{SabachSoker2014} estimated that the event rate of all routes of WD-NS reverse evolution is $\approx 3-5 \%$ of the CCSN rate, and about half end in fatal CEE, while about half of the cases end with a NS-WD binary system. I therefore estimate that cases of reverse evolution with fatal CEE contribute $\approx 2 \%$ of all CCSNe. 
 
The ILOT properties \citep{SabachSoker2014} are similar to those of ILOTs of SNe Ia progenitors. 

The initially more massive star transfers mass to the primary star (initially less massive star). For that, the WD remnant of the initially more massive star is unlikely to have a closer companion, and the triple stellar evolutionary channel that I study here is not relevant. 

\section{Summary}
\label{sec:summary}

I constructed a class of supernovae and progenitors of supernovae that result from the fatal CEE. 
I listed the seven members of the class and their properties in Table \ref{tab:Table1}. {{{{ In Fig. \ref{fig:Schematic} I schematically drew the seven scenarios. }}}}
Although each of these scenarios has been studied and discussed in the past, to the best of my knowledge no study grouped them into one class. 

Grouping the seven members of the class together into one class reveals the common evolutionary phases and properties they share (section \ref{sec:Properties}), and the differences between them as I discussed in section \ref{sec:classmembers} and listed in Table \ref{tab:Table1}. The main general achievement of the study is in presenting the potential of combining the two research areas of supernovae and CEE, each having many open questions, to shed light on each other. 

Specifically, I can summarise the main points of this study as follows
\begin{enumerate}
\item The notion that in many cases the CEE ends with the companion reaching the core might explain a large fraction of SNe Ia, and account for a non-negligible fraction of exploding massive stars, i.e., all CEJSNe and $\approx 6-10 \%$ of CCSNe. In the opposite direction, the supernovae that I discussed here, most of which are rare, can teach us about the conditions for fatal CEE.   
\item The interaction of the companion with the core might either destroy the core when the companion is a NS or a BH, destroy the companion when it is a main sequence star, or the two can merge to form a larger core in cases where the companion is a WD or a core of another giant. I argued that each one of these three possibilities might lead to supernovae and/or to the formation of  supernova progenitors. A detailed study of these supernovae might further shed light on the fate of the companion-core encounter in the CEE.  
\item Jets are involved in all the cases studied here. Jets might shape the CSM during the early CEE (in all cases beside the CEE of two giants), might facilitate the removal of the common envelope (for NS/BH/main sequence companions), and might power the explosion of CEJSN and might be the main powering mechanism of the CCSNe studied here. In these CCSNe the merger is likely to spin-up the core such that it has a non-negligible rotation when it finally collapses. A large angular momentum eases the formation of jets by the newly born NS. Some of the scenarios that I studied here, in particular CEJSNe, make a nice connection between jets during the CEE phase and jets at explosion. The comparison of the morphologies of CCSN remnants with morphologies of planetary nebulae that are shaped by CEE binaries can teach us about the role of jets in both the CEE and in the explosion of CCSN \citep{BearSoker2017Ears, Bearetal2017}. I encourage a similar comparison now using the CSM properties of some of the supernovae that I discussed here. 
\item Very rare cases are those when the companion is in a tight binary system. The spiralling-in of a tight binary system inside the common envelope can lead to even more peculiar properties as I listed in the last row of Table \ref{tab:Table1}.  
\end{enumerate}
 
With more sky coverage and larger surveys, more peculiar supernovae will be detected in the near future and with better exploration of their properties. This study raises the possibility that a large fraction of peculiar and rare supernovae result from the fatal CEE. Theoretical understanding of these supernovae requires us to better understand the fate of the CEE. In turn, the supernovae and CEE can teach us one about the other. 

I thank Efrat Sabach, Avishai Gilkis and Amit Kashi for helpful comments.  
{{{{ I thank two anonymous referees for very detailed and useful comments. }}}} 
This research was supported by the E. and J. Bishop Research Fund at the Technion and by a grant from the Israel Science Foundation.


\label{lastpage}

\begin{thebibliography}{}

\bibitem[Akras et al.(2015)]{Akarasetal2015} S. Akras, P. Boumis, J. Meaburn, J. Alikakos, J., A. Lopez, and D. R. Goncalves, \mnras, 452, 2911 (2015)
  
\bibitem[Andrews \& Smith(2018)]{AndrewsSmith2018} J. E. Andrews, and N. Smith, \mnras, 477, 74 (2018) 

\bibitem[Antoni et al.(2019)]{Antonietal2019} A. Antoni, M. MacLeod, and  E. Ramirez-Ruiz, arXiv:1901.07572 (2019)

\bibitem[Arcavi et al.(2017)]{Arcavietal2017} I. Arcavi, D. A. Howell, D. Kasen, D., et al., \nat, 551, 210 (2017)

\bibitem[Armitage \& Livio(2000)]{ArmitageLivio2000} P.J. Armitage, and M. Livio, \apj, 532, 540 (2000) 

\bibitem[Bear \& Soker(2017a)]{BearSoker2017} E. Bear, and N. Soker, \apjl, 837, L10 (2017a) 

\bibitem[Bear \& Soker(2017b)]{BearSoker2017Ears} E. Bear, and N. Soker, \mnras, 468, 140 (2017b) 

\bibitem[Bear et al.(2017)]{Bearetal2017} E. Bear, A. Grichener, and N. Soker, \mnras, 472, 1770 (2017)

\bibitem[Blackman \& Lucchini(2014)]{BlackmanLucchini2014} E. G. Blackman, and S. Lucchini, \mnras, 440, L16 (2014) 

\bibitem[Boffin et al.(2012)]{Boffinetal2012} H. M. J. Boffin, B. Miszalski, T. Rauch, D. Jones, R. L. M. Corradi, R. Napiwotzki, A. C. Day-Jones, and J. K\"oppen, Science, 338, 773 (2012) 

\bibitem[Bravo(2019)]{Bravo2019} {{{{ E. Bravo, arXiv:1903.08344  (2019) }}}}

\bibitem[Canals et al.(2018)]{Canalsetal2018} P. Canals, S. Torres, and N. Soker, \mnras, 480, 4519 (2018)

\bibitem[Chamandy et al.(2018)]{Chamandyetal2018} L. Chamandy, A. Frank, E. G. Blackman, et al., \mnras, 480, 1898 (2018) 

\bibitem[Chamandy et al.(2019)]{Chamandyetal2019} L. Chamandy, Y. Tu, E. G. Blackman, et al., arXiv:1812.1119 (2019) 

\bibitem[Chen et al.(2016)]{Chenetal2016} Z. Chen, J. Nordhaus, A. Frank, E. G. Blackman, and B. Balick, \mnras, 460, 4182 (2016)

\bibitem[Chevalier(2012)]{Chevalier2012} R. A. Chevalier, \apjl, 752, L2 (2012) 

\bibitem[Chevalier \& Soker(1989)]{ChevalierSoker1989} R. A. Chevalier, and N. Soker, \apj, 341, 867 (1989) 

\bibitem[Chugai(2018)]{Chugai2018} N. N. Chugai, Astronomy Letters, 44, 370 (2018) 

\bibitem[Claeys et al.(2011)]{Claeysetal2011} J. S. W. Claeys, S. E. de Mink, O. R. Pols, J. J. Eldridge, and M. Baes, M., \aap, 528, A131 (2011) 

\bibitem[Corradi et al.(2011)]{Corradietal2011aa} R. L. M. Corradi, B. Balick, B., and M. Santander-Garc{\'{\i}}a, \aap, 529, A43 (2011)

\bibitem[De Marco \& Izzard(2017)]{DeMarcoIzzard2017} O. De Marco, O., and R. G. Izzard, \pasa, 34, e001: (2017)

\bibitem[de Mink et al.(2014)]{DeMinketal2014} S. E. de Mink, H. Sana, N. Langer, R. G. Izzard, and F. R. N. Schneider, \apj, 782, 7 (2014) 

\bibitem[Dessart(2018)]{Dessart2018} L. Dessart, \aap, 610, L10 (2018) 

\bibitem[Dilday et al.(2012)]{Dildayetal2012} B. Dilday, D. A. Howell, S. B. Cenko, et al., Science, 337, 942 (2012) 
 
\bibitem[Dopita et al.(2018)]{Dopitaetal2018} M. A. Dopita, A. Ali, A. I. Karakas, D. Goldman, M. A. Amer, and R. S. Sutherland, \mnras, 475, 424  (2018) 

\bibitem[Eldridge et al.(2018)]{Eldridgeetal2018} J. J. Eldridge, L. Xiao, E. R. Stanway, N. Rodrigues, and N. Y. Guo, \pasa, 35, 49 (2018)

\bibitem[Fang et al.(2019)]{Fangetal2019} K. Fang, B. D. Metzger, K. Murase, I. Bartos, and K. Kotera, arXiv:1812.11673 (2019)

\bibitem[Gal-Yam(2019)]{GalYam2019} A. Gal-Yam, arXiv:1812.01428 (2019)

\bibitem[Gilkis et al.(2016)]{Gilkisetal2016} A. Gilkis, N. Soker, and O. Papish, \apj, 826, 178 (2016)

\bibitem[Ginat et al.(2019)]{Ginatetal2019} {{{{ Y. B. Ginat, H. B. Perets, E. Grishin, and V. Desjacques, preprint (2019) }}}}

\bibitem[Glanz \& Perets(2018)]{GlanzPerets2018} H. Glanz, and H. B. Perets, \mnras, 478, L12 (2018) 

\bibitem[Grichener \& Soker(2019)]{GrichenerSoker2019} A. Grichener, and N. Soker, arXiv:1810.03889 (2019)

\bibitem[Hachisu et al.(1999)]{Hachisuetal1999} I. Hachisu, M., Kato, and K. Nomoto, \apj, 522, 487 (1999)

\bibitem[Hillel et al.(2017)]{Hilleletal2017} S. Hillel, R., Schreier, and N. Soker, \mnras, 471, 3456 (2017)

\bibitem[Houck \& Chevalier(1991)]{HouckChevalier1991} J. C. Houck, and R. A. Chevalier, \apj, 376, 234 (1991)

\bibitem[Iaconi et al.(2017)]{Iaconietal2017} R. Iaconi, T. Reichardt, J. Staff, O. De Marco, J. C. Passy, D. Price, J. Wurster, and F. Herwig, \mnras, 464, 4028 (2017)

\bibitem[Iaconi \& De Marco(2019)]{IaconiDeMarco2019} R. Iaconi, and O. De Marco, arXiv:1902.02039  (2019) 

\bibitem[Iaconi et al.(2018)]{Iaconietal2018} R. Iaconi, O. De Marco, J. C. Passy, and J. Staff, mnras, 477, 2349 (2018)

\bibitem[Ivanova(2017)]{Ivanova2017} N. Ivanova, The Lives and Death-Throes of Massive Stars, 329, 199 (2017)

\bibitem[Ivanova et al.(2013a)]{Ivanovaetal2013Science} N. Ivanova, S. Justham, J. L. Avendano Nandez, and J. C. Lombardi, Science, 339, 433 (2013a) 

\bibitem[Ivanova et al.(2013b)]{Ivanovaetal2013} N. Ivanova, S. Justham, S., X. Chen, et al., \aapr, 21, 59 (2013b)

\bibitem[Jencson et al.(2019)]{Jencsonetal2019} {{{{ J. E. Jencson, M. M. Kasliwal, S. M. Adams, et al., arXiv:1901.00871 (2019) }}}}

\bibitem[Jones(2019)]{Jones2019} D. Jones, arXiv:1806.08244 (2019) 
 
\bibitem[Jones \& Boffin(2017)]{JonesBoffin2017} D. Jones, and H. M. J. Boffin, Nature Astronomy, 1, 0117  (2017)	

\bibitem[Jones et al.(2018)]{Jonesetal2018} D. Jones, H. M. J. Boffin, P. Sowicka, B. Miszalski, P. Rodriguez-Gil, M. Santander-Garcia, R. L. M. Corradi, \mnras, 482, L75 (2018) 

\bibitem[Justham et al.(2014)]{Justhametal2014} S. Justham, P. Podsiadlowski, and J. S. Vink, \apj, 796, 121 (2014)

\bibitem[Kashi \& Soker(2016)]{KashiSoker2016RAA} {{{{ A. Kashi, and N. Soker, Research in Astronomy and Astrophysics, 16, 99 (2016) }}}}

\bibitem[Kashi \& Soker(2018)]{KashiSoker2018} A. Kashi, and N. Soker, \apj, 858, 117 (2018)  

\bibitem[Kotak \& Vink(2006)]{KotakVink2006} R. Kotak, and J. S. Vink, \aap, 460, L5 (2006)

\bibitem[Kuin et al.(2019)]{Kuinetal2019} N. P. M. Kuin, K. Wu, S. Oates, et al., arXiv:1808.08492 (2019)

\bibitem[Kuruwita et al.(2016)]{Kuruwitaetal2016}  R. L. Kuruwita, J. Staff, and O. De Marco, \mnras, 461, 486 (2016)

\bibitem[Livio \& Mazzali(2018)]{LivioMazzali2018} M. Livio, and P. Mazzali, Physics Reports, 736, 1 (2018)

\bibitem[Livio \& Riess(2003)]{LivioRiess2003} M. Livio, and A. G. Riess, \apjl, 594, L93  (2003)

\bibitem[Lohev et al.(2019)]{Lohevetal2019} N. Lohev, E. Sabach, E., A. Gilkis, and N. Soker, in preparation (2019)

\bibitem[Lyutikov \& Toonen(2019)]{LyutikovToonen2019} M. Lyutikov, and S. Toonen, arXiv:1812.07569  (2019) 

\bibitem[MacLeod et al.(2018a)]{MacLeodetal2018a} M. MacLeod, E. C. Ostriker, and J. M. Stone, \apj, 863, 5 (2018a)

\bibitem[MacLeod et al.(2018b)]{MacLeodetal2018b} M. MacLeod, E. C. Ostriker, and J. M.  Stone, \apj, 868, 136 (2018b)

\bibitem[Maoz et al.(2014)]{Maozetal2014} D. Maoz, F. Mannucci, and G. Nelemans, \araa, 52, 107 (2014)

\bibitem[Menon \& Heger(2017)]{MenonHeger2017} A. Menon, and A. Heger, \mnras, 469, 4649 (2017) 

\bibitem[Margutti et al.(2019)]{Marguttietal2019} R. Margutti, D. D. Metzger, R. Chornock, et al., arXiv:1810.10720 (2019)

\bibitem[Michaely \& Perets(2019)]{MichaelyPerets2019} E. Michaely, and H. B. Perets, \mnras, 484, 4711 (2019) 

\bibitem[Miszalski et al.(2011)]{Miszalskietal2011} B. Miszalski, R. L. M. Corradi, H. M. J. Boffin, D. Jones, L. Sabin, M. Santander-Garcia, P. Rodriguez-Gil, P., and M. M. Rubio-Diez, \mnras, 413, 1264: ETHOS 1 (2011)
   
\bibitem[Miszalski et al.(2018)]{Miszalskietal2018} B. Miszalski, R. Manick, J. Miko{\l}ajewska, H. Van Winckel, Hand K. I{\l}kiewicz, \pasa, 35, e027 (2018)
 
\bibitem[Nandez \& Ivanova(2016)]{NandezIvanova2016} J. L. A. Nandez, and N. Ivanova, \mnras, 460, 3992 (2016)

\bibitem[Nandez et al.(2014)]{Nandezetal2014} J. L. A. Nandez, N. Ivanova, and J. C. Jr. Lombardi, \apj, 786, 39 (2014) 

\bibitem[Nazin \& Postnov(1997)]{Nazinetal1997} {{{{ S. N. Nazin, S.~N., amd K. A. Postnov, Astronomy Letters, 23, 139 (1997) }}}}

\bibitem[Nomoto et al.(1995)]{Nomotoetal1995} K. T. Nomoto, K. Iwamoto, and T. Suzuki, Physics Reports, 256, 173 (1995)

\bibitem[Ohlmann et al.(2016)]{Ohlmannetal2016} S. T. Ohlmann, F. K. R{\"o}pke, R. Pakmor, and V. Springel, \apjl, 816, L9 (2016)

\bibitem[Papish et al.(2015)]{Papishetal2015} O. Papish, N. Soker, and I. Bukay, \mnras, 449, 288 (2015)

\bibitem[Passy et al.(2012)]{Passyetal2012} J. C. Passy, O. De Marco, C. L. Fryer, et al., \apj, 744, 52 (2012) 

\bibitem[Pastorello et al.(2012)]{Pastorelloetal2012} A. Pastorello, M. L. Pumo, H. Navasardyan, et al., \aap, 537, A141 (2012) 

\bibitem[Perley et al.(2019)]{Perleyetal2019} D. A. Perley, P. A. Mazzali, L. Yan, et al., \mnras, \mnras, 484, 1031 (2019)

\bibitem[Podsiadlowski et al.(1990)]{Podsiadlowskietal1990} P. Podsiadlowski, P. C. Joss, and S. Rappaport, \aap, 227, L9 (1990) 

\bibitem[Quataert et al.(2019)]{Quataertetal2019} E. Quataert, D. Lecoanet, and E. R. Coughlin, ArXiv e-prints , arXiv:1811.12427 (2019) 

\bibitem[Reichardt et al.(2019)]{Reichardtetal2019} T. A. Reichardt, O. De Marco, R. Iaconi, C. A. Tout, and D. J. Price, \mnras, 484, 631 (2019)

\bibitem[Retter \& Marom(2003)]{RetterMarom2003} A. Retter, and A. Marom, \mnras, 345, L25 (2003)

\bibitem[Ricker \& Taam(2012)]{RickerTaam2012} P. M. Ricker, and R. E. Taam, \apj, 746, 74 (2012) 

\bibitem[Ruiz-Lapuente(2019)]{RuizLapuente2019} P. Ruiz-Lapuente, arXiv e-prints , arXiv:1812.04977 (2019)

\bibitem[Sabach \& Soker(2014)]{SabachSoker2014} E. Sabach, and N. Soker, \mnras, 439, 954 (2014)

\bibitem[Sabach \& Soker(2015)]{SabachSoker2015} E. Sabach, and N. Soker, \mnras, 450, 1716 (2015)

\bibitem[Sahai et al.(2016)]{Sahaietal2016} R. Sahai, S. Scibelli, and M. R. Morris, \apj, 827, 92 (2016) 
 
\bibitem[Sahai et al.(2017)]{Sahaietal2017} R. Sahai, W. H. T. Vlemmings, and L.-{\AA} Nyman,  \apj, 841, 110 (2017)

\bibitem[Seitenzahl et al.(2013)]{Seitenzahletal2013} I. R. Seitenzahl, G. Cescutti, F. K. R{\"o}pke, A. J. Ruiter, and R. Pakmor, \aap, 559, L5 (2013) 

\bibitem[Shiber(2018)]{Shiber2018} S. Shiber, Galaxies, 6, 96 (2018)

\bibitem[Shiber et al.(2017)]{Shiberetal2017} S. Shiber, A. Kashi, and N. Soker, \mnras, 465, L54 (2017)

\bibitem[Shiber et al.(2019)]{Shiberetal2019} S. Shiber, S., R. Iaconi, O. De Marco, and N. Soker, arXiv:1902.03931 (2019)

\bibitem[Shivvers et al.(2017)]{Shivversetal2017} I. Shivvers, M. Modjaz, W. Zheng, et al., \pasp, 129, 054201 (2017) 

\bibitem[Smith(2014)]{SmithN2014} N. Smith, \araa, 52, 487 (2014)

\bibitem[Soker(2016a)]{Soker2016messy} N. Soker, \mnras, 455, 1584 (2016a)

\bibitem[Soker(2016b)]{Soker2016Rev} N. Soker, \nar, 75, 1 (2016b)

\bibitem[Soker(2017a)]{Soker2017RAA} N. Soker, Research in Astronomy and Astrophysics, 17, 113 (2017a)
  
\bibitem[Soker(2017b)]{Soker2017IIb}  N. Soker, \mnras, 470, L102 (2017b)

\bibitem[Soker(2018)]{Soker2018Rev} N. Soker, Science China Physics, Mechanics, and Astronomy, 61, 49502 (2018)

\bibitem[Soker(2019a)]{Soker2019CEexit} N. Soker, \mnras, 483, 5020 (2019a) 

\bibitem[Soker(2019b)]{Soker2019SASI} N. Soker, Research in Astronomy and Astrophysics; arXiv:1810.09074 (2019b)

\bibitem[Soker \& Gilkis(2018)]{SokerGilkis2018} N. Soker, and A. Gilkis, \mnras, 475, 1198 (2018)

\bibitem[Soker et al.(2019)]{Sokeretal2019AT2018cow} N. Soker, A. Grichener, and A. Gilkis, \mnras, 484, 49726 (2019)

\bibitem[Soker \& Kashi(2012)]{SokerKashi2012} N. Soker, and A. Kashi, \apj, 746, 100 (2012) 
 
\bibitem[Soker \& Kashi(2016)]{SokerKashi2016} N. Soker, and A. Kashi, A., \mnras, 462, 217 (2016)

\bibitem[Soker et al.(2013)]{Sokeretal2013PTF11kx} N. Soker, A. Kashi, E. Garc{\'{\i}}a-Berro, S. Torres, and J. Camacho, \mnras, 431, 1541 (2013) 

\bibitem[Sowicka et al.(2017)]{Sowickaetal2017} P. Sowicka, D. Jones, R. L. M. Corradi, et al., \mnras, 471, 3529 (2017)

\bibitem[Sravan et al.(2018)]{Sravanetal2018} N. Sravan, P. Marchant, and V. Kalogera, V., \& Margutti, R.\ 2018 \apj, 852, L17 (2018)

\bibitem[Sravan et al.(2019)]{Sravanetal2019} N. Sravan, P. Marchant, and V. Kalogera, V.\ 2019, arXiv:1808.07580 (2019) 

\bibitem[Staff et al.(2016)]{Staffetal2016MN8} J. E. Staff, O. De Marco,  P. Wood, P. Galaviz, and J. C. Passy, \mnras, 458, 832 (2016)

\bibitem[Tauris et al.(2015)]{Taurisetal2015} T. M. Tauris, N. Langer, and P. Podsiadlowski, \mnras, 451, 2123 (2015)
 
\bibitem[Tocknell et al.(2014)]{Tocknelletal2014} J. Tocknell, O. De Marco, and M. Wardle, \mnras, 439, 2014 (2014)

\bibitem[Tsebrenko \& Soker(2015)]{TsebrenkoSoker2015} D. Tsebrenko, and N. Soker, \mnras, 447, 2568 (2015)

\bibitem[Tylenda et al.(2011)]{Tylendaetal2011} R. Tylenda, M. Hajduk, T. Kami{\'n}ski, T., et al., \aap, 528, A114 (2011) 

\bibitem[Tylenda et al.(2013)]{Tylendaetal2013} R. Tylenda, T. Kami{\'n}ski, A. Udalski, et al., \aap, 555, A16 (2013) 
 
\bibitem[Urushibata et al.(2018)]{Urushibataetal2018} T. Urushibata, K. Takahashi, H. Umeda, and T. Yoshida, \mnras, 473, L101 (2018)

\bibitem[van den Heuvel et al.(2017)]{vandenHeuveletal2017} E. P. J. van den Heuvel, S. F. Portegies Zwart, and S. E. de Mink, \mnras, 471, 4256 (2017)

\bibitem[Vel{\'a}zquez et al.(2011)]{Velazquezetal2011ApJ} P. F. Vel{\'a}zquez, W. Steffen, A. C. Raga, S. Haro-Corzo, A. Esquivel, J. Canto, and A. Riera, \apj, 734, 57 (2011)

\bibitem[Vigna-G{\'o}mez et al.(2019)]{VignaGomezetal2019} {{{{ A. Vigna-G{\'o}mez, S. Justham, I. Mandel, S. E. de Mink, and P. Podsiadlowski, arXiv:1903.02135  (2019) }}}}

\bibitem[Vigna-G{\'o}mez et al.(2018)]{VignaGomezetal2018} A. Vigna-G{\'o}mez, C. J. Neijssel, S. Stevenson, et al., \mnras, 481, 4009 (2018)

\bibitem[Wang(2018)]{Wang2018} B. Wang, Research in Astronomy and Astrophysics, 18, 49 (2018)

\bibitem[Wang et al.(2018)]{Wangetal2018} L. J. Wang, X. F. Wang, S. Q. Wang, et al., \apj, 865, 95 (2018) 

\bibitem[Wilson \& Nordhaus(2019)]{WilsonNordhaus2019} E. C. Wilson, and J. Nordhaus, arXiv:1811.03161 (2019)

\bibitem[Woosley(2018)]{Woosley2018} S. E. Woosley, \apj, 863, 105 (2018)

\bibitem[Yoon(2015)]{Yoon2015} S. C. Yoon, \pasa, 32, e015 (2015) 

\bibitem[Yoon(2017)]{Yoon2017} S. C. Yoon, \mnras, 470, 3970 (2017)

\bibitem[Young et al.(2006)]{Youngetal2006} P. A. Young, C. L. Fryer, A. Hungerford, et al., \apj, 640, 891 (2006)

\bibitem[Zapartas et al.(2017)]{Zapartasetal2017} E. Zapartas, S. E. de Mink, R. G. Izzard, et al., \aap, 601, A29 (2017)

\end{thebibliography}
\end{document}